\documentclass{iopart}
\pdfoutput=1
\usepackage{graphicx}
\usepackage{cite}
\usepackage{comment}
\usepackage{epsfig}
\usepackage{bm}
\usepackage{color}
\usepackage{times}
\usepackage{braket}
\usepackage{cancel}
\usepackage[colorlinks,bookmarks=false,citecolor=blue,linkcolor=red,urlcolor=blue]{hyperref}
\hypersetup{%
  colorlinks = true,
  linkcolor  = black
} 

\def\doi{http://dx.doi.org/}
\usepackage{tikz}
\usetikzlibrary{arrows}
\usetikzlibrary{decorations.pathreplacing}

\usepackage[colorlinks,bookmarks=false,citecolor=blue,linkcolor=red,urlcolor=blue]{hyperref}
\hypersetup{%
  colorlinks = true,
  linkcolor  = black
} 

\def\doi{http://dx.doi.org/}

\newcommand{\be}{\begin{equation}}
\newcommand{\ee}{\end{equation}}
\newcommand{\bea}{\begin{eqnarray}}
\newcommand{\eea}{\end{eqnarray}}

\begin{document}

\title{The density profile of the six vertex model with domain wall boundary conditions}
\author{I. Lyberg$^1$, V. Korepin $^{2}$ and J. Viti$^{1,3}$}
\address{$^1$Instituto Internacional de Fisica, UFRN,  Campos Universit\' ario, Lagoa Nova  59078-970 Natal, Brazil}
\address{$^2$C.N. Yang Institute for Theoretical Physics, Stony Brook University, Stony Brook, USA}
\address{$^3$ Escola de Ciencia e Tecnologia, UFRN,  Campos Universit\' ario, Lagoa Nova  59078-970 Natal, Brazil\footnote{Email for correspondence: jacopo.viti@ect.ufrn.br}}

\begin{abstract}
We  study numerically the density profile in the six-vertex model with domain wall boundary conditions.
Using a Monte Carlo algorithm originally proposed by Allison and Reshetikhin we numerically evaluate the inhomogeneous density profiles
in the disordered and antiferromagnetic regimes where frozen corners appear. 
At the free fermion point we present an exact finite-size formula for the density on the horizontal edges that relies on the imaginary time transfer matrix
approach. In all cases where  exact analytic forms for the density and the arctic curves  are known the numerical method 
shows perfect agreement with them. This also suggests the possibility of its use for accurate quantitative purposes.
\end{abstract}
\maketitle

\frenchspacing

\section{Introduction}
The six vertex model is a well-known and studied model in statistical mechanics. It was proposed long ago by Linus Pauling~\cite{P35} to model the
ice-water transition and has become an important example of the crucial role that boundary conditions may play in systems that satisfy local constraints such as
the ice-rule. The model was solved by Elliot Lieb~\cite{Lieb} for periodic boundary conditions with transfer matrix techniques,
leading to the famous phase diagram as a function of the control parameter $\Delta$; see also~\cite{Baxter}. The influence of the boundary conditions is however so
severe~\cite{zj_bc} that it is legitimate to ask which among them lead to the free energy found by Lieb and which do not.
In particular, in~\cite{kzj} it was proved that with so called \textit{Domain Wall Boundary Conditions (DWBC)},
the extensive part of the free energy is not the same as with periodic boundary conditions. The DWBC were originally introduced 
in~\cite{Korepin} to analyze  the norm of Bethe Ansatz eigenstates; a pedagogical discussion is contained in~\cite{KBI}.
The authors of \cite{kzj} used a determinant
representation of the partition function~\cite{Izergin, ICK} in order to derive Toda differential equations, whose
asymptotic solutions gave the extensive part of the free energy.  
Subleading corrections to the free energy as a function of the system size were later determined in~\cite{Bleher} in all the phases of the model. 

Recent interest in
the six vertex model with DWBC was also triggered by the phenomenon of  phase separation~\cite{kzj, zj2000, syljuaasen} and its manifestation
through the appearance of an \textit{arctic curve} or in a more general jargon a limiting shape~\cite{KO, KOS}.

The
arctic curve is the boundary between a disordered region that lies in its interior and a completely frozen domain in its exterior.
At the free fermion point $\Delta=0$, the arctic curve is an ellipse~\cite{Propp}, as can be argued from
the mapping between the model and dimer coverings on the Aztec diamond~\cite{FS, ZJ_rev}. Moreover when $a=b$ and $c=\sqrt{2}a$,
see formula~(\ref{delta}) in Sec.~\ref{sec_1}, the ellipse is actually a circle: the  arctic circle. The density profile on the vertical or horizontal edges
studied in this
paper and defined later in Sec.~\ref{sec_1} is equivalent to the polarization considered  in~\cite{BPZ_boundary, syljuaasen}.
When $\Delta=0$ it can  be derived exactly
by introducing a diagonal-to-diagonal transfer matrix~\cite{viti} and exploiting free fermionic techniques in imaginary time that are very close to usual
real time quantum mechanical methods.

 Away from the free fermion point, exact expressions are known for the arctic curve, conjectured in the disordered~\cite{PC}  ($|\Delta|\leq 1$) and
 antiferromagnetic~\cite{PCZ} ($\Delta<-1$) regimes; see also~\cite{CS}. However no results  are available for the density.  Attempts of deriving it as a
 solution  of a partial differential equation  that  generalizes to arbitrary values of $\Delta$
or includes other type of boundary conditions are in progress~\cite{Abanov, Reshetikhin2}.

The problem of determining the  density  of the six vertex model with DWBC on the whole phase diagram is intrinsically difficult.
Therefore the implementation of a reliable numerical method is of clear relevance. An important step in this direction was taken by Allison and Reshetikhin in 
2005 \cite{reshetikhin}, who proposed a Monte Carlo
algorithm to generate a typical state. Such a state could then be compared with
the known phase diagram for periodic boundary conditions.  The authors obtained  typical states for all phases.  In particular
it was found that in the antiferromagnetic regime ($\Delta < -1$) ,
three phases can coexist. Earlier numerical studies were done in ~\cite{syljuaasen, barkema}, where different Monte Carlo algorithms were used, and recently in~\cite{cugliandolo}. In this paper we reconsider the Allison-Reshetikhin algorithm  as a tool to obtain  numerical estimates of the density profile in the disordered  and  antiferromagnetic
regimes, where essentially no exact results are known.

 We test the algorithm at $\Delta=0$, relying on explicit  formulas in~\cite{viti}.
In particular we  present an exact finite size expression for the density on the horizontal edges at the free fermion point that shows excellent agreement with the numerics.
This is an original and strong test of the reliability of the numerical method that also suggests the possibility of its use  to obtain accurate quantitative results and not only qualitative information on the phase diagram.

The paper is organized as follows. In Sec.~\ref{sec_1} we review the imaginary time approach to the six vertex model with DWBC and present an exact finite size formula for the density at
$\Delta=0$. In Sec.~3 and Sec.~4 we show  numerical results for the density profile in the disordered and antiferromagnetic regimes. We also provide a comparison between the exact arctic curves conjectured in~\cite{PC, PCZ} and the their numerical estimate both in the disordered and anti-ferromagentic regimes. Our conclusions and perspectives are gathered in
the final section. 

Finally we note that when $\Delta = 0$, the effect of the side-length of the lattice, $N$, has also been studied from a different point of view.
In~\cite{kj2000, kj, FS}, the fluctuations of the boundary of the arctic circle in the thermodynamic limit were shown to be given by a determinantal process known as the Airy process~\cite{PS}.
Suitably defined two-point correlation functions divided by a factor $N^{1/3}$ should be proportional to the Airy kernel~\cite{TW}. Unfortunately the numerical resolution is not 
good enough to make a quantitative study in this case.    

\section{The Six Vertex Model with DWBC as a fermionic model in imaginary time}
\label{sec_1}
\begin{figure}[ht]
\centering
\begin{tikzpicture}[scale=0.5]
\begin{scope}
\clip (0,0) rectangle (14.3cm,14.1cm); 

\draw[style=help lines,dashed,rotate=45] (-20,-20) grid[step=1cm] (20,20); 
\draw[very thick] (1.414*5,0)--(1.414*5-1.414/2,1.414/2);
\foreach \x in {0,1,...,4}{                           
    \foreach \y in {0}{                       
    \draw[very thick] (1.414*\x,\y)--(1.414*\x+1.414/2,\y+1.414/2) {};
    \draw[very thick] (1.414*\x,\y)--(1.414*\x-1.414/2,\y+1.414/2) {};
    }
\foreach \x in {0,1,...,4}{                           
    \foreach \y in {10}{                       
    \draw[very thick] (1.414*\x,1.414*\y)--(1.414*\x+1.414/2,1.414*\y-1.414/2) {};
    }
    }
\foreach \x in {1,...,5}{                           
    \foreach \y in {10}{                       
    \draw[very thick] (1.414*\x,1.414*\y)--(1.414*\x-1.414/2,1.414*\y-1.414/2) {};
    }
    }
\draw[thick](0,1.414*5)--(20,1.414*5);
\draw[thick](5*1.414,0)--(5*1.414,20*1.414);
\draw[gray, very thick, fill=gray, opacity=0.1](5*1.414,1*1.414)--(1*1.414,5*1.414)--(5*1.414,9*1.414)--(9*1.414,5*1.414)--cycle;
\foreach \n in {0,...,8}{                           
               \draw[very thick] (1.414+1.414/2*\n, 5*1.414-1.414/2*\n)--(0,5*1.1414-\n*1.414) {};
    
    }
\foreach \n in {0,...,5}{                           
               \draw[very thick] (5*1.414-1.414*\n, 0)--(0,5*1.414-\n*1.414) {};
    
    }    

}
\foreach \n in {0,...,8}{                           
               \draw[very thick] (1.414+1.414/2*\n, 5*1.414+1.414/2*\n)--(0,6*1.414+\n*1.414) {};
    
    }

\foreach \n in {0,...,5}{                           
               \draw[very thick] (5*1.414-1.414*\n, 10*1.414)--(0,5*1.414+\n*1.414) {};
    
    }    

\foreach \n in {0,...,5}{                           
               \draw[black!60]  (5*1.414+1.414*\n, 0)--(10*1.414,5*1.414-\n*1.414) {};
    
    }
   
\foreach \n in {0,...,8}{                           
               \draw[black!60] (5*1.414+1.414/2*\n, 1.414+1.414/2*\n)--(6*1.414+1.414*\n,0) {};
    
    }

\foreach \n in {0,...,8}{                           
               \draw[black!60] (5*1.414+1.414/2*\n, 9*1.414-1.414/2*\n)--(6*1.414+1.414*\n,10*1.414) {};
    
    }
\foreach \n in {0,...,5}{                           
               \draw[black!60] (5*1.414+1.414*\n, 10*1.414)--(10*1.414+1.414*\n,5*1.414) {};
    
    }   
    
 \draw[very thick, red,->](5*1.414,0*1.414)--(10*1.414,5*1.414);
  \draw[very thick, red,->](5.5*1.414,0*1.414)--(0.25*1.414,5.25*1.414);
  \foreach \n in {0,...,8}{                           
               \draw[red] (8.3085+1.414/2*\n, 0.8839+1.414/2*\n)--(7.955+1.414/2*\n,1.237+1.414/2*\n) {};
   
     }
\node[above, red] at (10*1.414,5*1.414) {$x$}; 
\node[above, red] at (0.1,5*1.414) {$y$};

\node[red] at (5.75*1.414,5.75*1.414) {$\bullet$};
\draw[red, dashed] (5.75*1.414,5.75*1.414)--(8.13173+1.414/2*5, 1.06066+1.414/2*5);
\draw[red, dashed] (5.75*1.414,5.75*1.414)--(3*1.414, 3*1.414);
\node[above, red] at (5.75*1.414+0.1,5.75*1.414) {$\langle\rho_h(x,y)\rangle$};
\foreach \n in {-9,...,9}{                           
               \draw[black!60] (5*1.414-0.2, 5*1.414+1.414/2*\n)--(5*1.414+0.2, 5*1.414+1.414/2*\n) {};
   
     }
\foreach \n in {-9,...,9}{                           
               \draw[black!60] (5*1.414+1.414/2*\n, 5*1.414-0.2)--(5*1.414+1.414/2*\n, 5*1.414+0.2) {};
   
     }     
\node[right] at (5*1.414,9.5*1.414) {$N$}; 
\node[right] at (5*1.414,0.5*1.414) {$-N$};
\draw[dashed] (0,1/4*1.414)--(10*1.414,1/4*1.414);
\draw[dashed] (0,9.75*1.414)--(10*1.414,9.75*1.414);
\end{scope}
\begin{scope}[xshift=15cm]
\draw[->, ultra thick](0,0)--(0,10*1.414);
\node[rotate=90] at(0.5,5*1.414) {$\rm{imaginary~~time:}~\textit{y}'$};
\end{scope}
\begin{scope}[yshift=-0.5cm]
\draw[->, ultra thick](0,0)--(10*1.414,0);
\node[->, below] at (5*1.414,0){$\rm{space:}~\textit{x}'$};
\end{scope}
\begin{scope}[xshift=-0.5cm]
\node[left] at(0,1/4*1.414) {$|B_i\rangle=|DW\rangle$};
\node[left] at(0,9.75*1.414) {$\langle B_f|=\langle DW|$};
\end{scope}
\end{tikzpicture}
\caption{The six-vertex model on a $N\times N$ square lattice with DWBC ($N=9$ in the figure)
is obtained from the discrete imaginary time evolution of fermionic trajectories
on a strip when
the initial and final state is the \textit{domain wall initial state}. The coordinate
system $(x,y)$ is defined as above
with $\langle \rho_h\rangle$ the density of fermions on the horizontal edges discussed in the main text.}
\label{fig_1}
\end{figure}
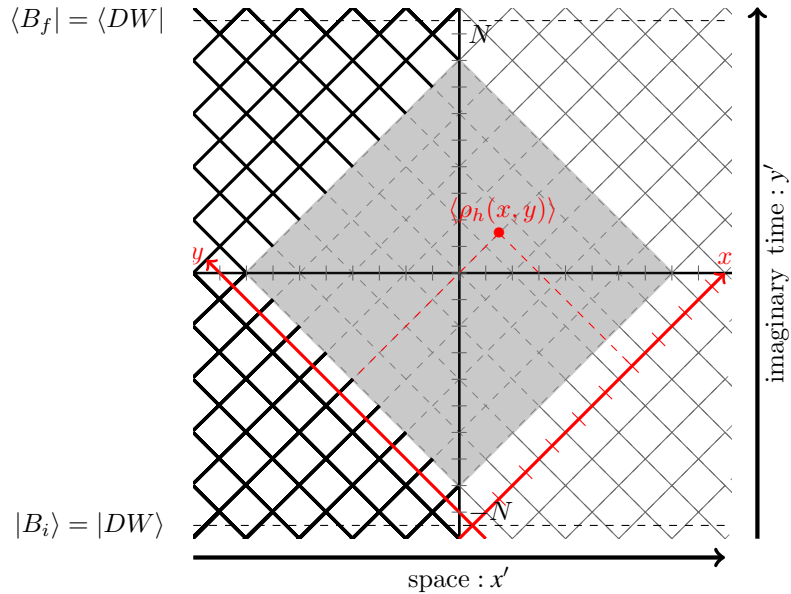
\begin{figure}[t]
\centering 
\begin{tikzpicture}[scale=0.6]
\draw[ultra thick](-1,-1)--(1,1);
\draw[ultra thick](1,-1)--(-1,1);
\node[below] at (0,-1) {$a_1$};
\draw [decorate,decoration={brace,amplitude=10pt},xshift=-1cm,yshift=1.5cm, rotate=-90, thick]
(0,-0.5) -- (0,5.0) node [black,midway,yshift=0.7cm] 
{\rm{weight $a$}};
\begin{scope}[xshift=3cm]
\draw(-1,-1)--(1,1);
\draw(1,-1)--(-1,1);
\node[below] at (0,-1) {$a_2$};
\end{scope}
\begin{scope}[xshift=6cm]
\draw(-1,-1)--(1,1);
\draw[ultra thick](1,-1)--(-1,1);
\node[below] at (0,-1) {$b_1$};
\draw [decorate,decoration={brace,amplitude=10pt},xshift=-1cm,yshift=1.5cm, rotate=-90, thick]
(0,-0.5) -- (0,5.0) node [black,midway,yshift=0.7cm] 
{\rm{weight $b$}};
\end{scope}
\begin{scope}[xshift=9cm]
\draw[ultra thick](-1,-1)--(1,1);
\draw(1,-1)--(-1,1);
\node[below] at (0,-1) {$b_2$};
\end{scope}
\begin{scope}[xshift=15cm]
\draw (-1,-1)--(0,0);
\draw [ultra thick](0,0)--(1,1);
\draw [ultra thick] (1,-1)--(0,0);
\draw (0,0)--(-1,1);
\node[below] at (0,-1) {$c_2$};
\end{scope}

\begin{scope}[xshift=12cm]
\draw [ultra thick](-1,-1)--(0,0);
\draw (0,0)--(1,1);
\draw (1,-1)--(0,0);
\draw [ultra thick] (0,0)--(-1,1);
\node[below] at (0,-1) {$c_1$};
\draw [decorate,decoration={brace,amplitude=10pt},xshift=-1cm,yshift=1.5cm, rotate=-90, thick]
(0,-0.5) -- (0,5.0) node [black,midway,yshift=0.7cm] 
{\rm{weight $c$}};
\end{scope}
\end{tikzpicture}
\caption{When one or two fermionic particles denoted by thick lines meet at one vertex of the lattice, we have six possible vertices. We assign to each of them the weights $a,b$ and $c$ as above. Fermionic trajectories
on the lattice are then in one-to-one correspondence with allowed configurations of the six vertex model.}
\label{fig_2}
\end{figure}
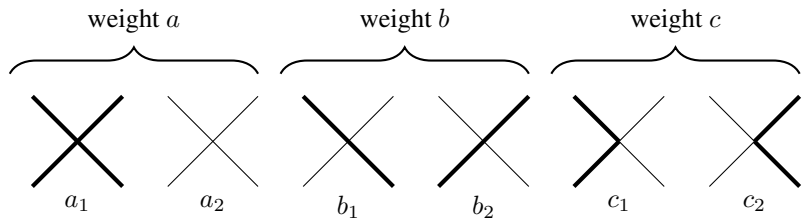

We briefly review  the imaginary
time formalism discussed in~\cite{viti} to the six vertex model with DWBC.
Consider an
infinitely long strip embedded into  a square
lattice, as depicted in Fig.~\ref{fig_1}.
The horizontal coordinate $x'$ is conventionally called \textit{space} 
and the vertical $y'$ \textit{imaginary time};  both are discrete. The total height of the strip in
Fig.~\ref{fig_1} is $2(N+1)$. $N$ is taken odd for simplicity.
 We assume that  each edge of the  lattice
 can host a fermionic particle. We mark the edge with a thick line if 
 the particle is  present, and with a thin line if it is absent. More precisely, to any edge is associated the Hilbert space $\mathbf{C}^2$ on which the fermionic creation and annihilation operators $c_{x'}$ and $c_{x'}^{\dagger}$ act; pictorially $|0\rangle$ is a thin edge and $c^{\dagger}_{x'}|0\rangle$ a thick one.
Notice also that fermionic spatial and temporal
 coordinates are  semi-integer. Fermionic trajectories cannot cross and evolve in discrete imaginary time,
 according to the following dynamical rule.
 When one or two fermions meet at one vertex, six possible different processes may take place that we label
  as $a_1,a_2,b_1,b_2,c_1,c_2$; see Fig.~\ref{fig_2}. We then assign three 
 weights $a,b$ and $c$ to any one of those, according to Fig.~\ref{fig_2}.
These processes are the matrix elements of the six-vertex $R$-matrix~\cite{resh_rev} $R_{x',x'+1}$. The $R$-matrix acts 
on the four dimensional Hilbert space $\mathbf C^2\otimes \mathbf C^2$ of two neighbouring fermionic
particles located at spatial semi-integer coordinates $x'$ and $x'+1$.
Since fermionic trajectories cannot cross, they are in one-to-one correspondence with a particular
edge configuration of the six vertex model. This is of course equivalent to the usual line interpretation, see for example~\cite{BPZ_boundary}.
Given the $R$-matrix we can construct two transfer matrices
implementing evolution in imaginary
time by simply taking $T_{e/o}=\prod_{x'\in 2 \mathbf Z\pm 1/2} R_{x',x'+1}$. The elementary evolution
operator can be, for example, the double row-to-row  transfer matrix $\tilde{T}=T_eT_o$. In the imaginary time
formalism boundary conditions are imposed \textit{only} on the edges at the top and bottom of the strip. We fix weights of the edges at
constant $y'=\pm(N+\frac{1}{2})$, see the two dashed lines in Fig. ~\ref{fig_1}.
In the fermionic language this is equivalent to specifying two particular boundary states
$|B_i\rangle$ and $|B_f\rangle$ as initial and final states for the
discrete fermionic imaginary-time evolution. The partition function for the lattice
model is then the amplitude
\begin{equation}
 Z=\langle B_f|\tilde{T}^N T_e|B_i\rangle=\langle B_f| T_e^{1/2}T^{2N} T_{e}^{1/2}|B_i\rangle,
\end{equation}
with $T^2=T_e^{1/2}T_oT_e^{1/2}$ an hermitian operator. The formalism can be adapted to compute correlation functions rather straightforwardly. Perhaps the most relevant example is the fermion density $\langle \rho_{\rm{even}}(x',y')\rangle =\langle c^{\dagger}_{x',y'}c_{x',y'}\rangle$
 at the point $(x',y')$ with  $x'-y' \in 2\mathbf Z$. The fermion density is given by the equation
\begin{eqnarray} \nonumber
\hspace*{-2cm} & &\langle \rho_{\rm{even}}(x',y') \rangle =  \\
& & Z^{-1}\langle B_f| T_e^{1/2}T^{2\left(\frac{N\mp1/2-y'}{2}\right)}T_e^{1/2} c^{\dagger}_{x'}c_{x'} T_{e}^{-1/2}T^{2\left(\frac{N\mp1/2+y'}{2}\right)}T_{e}^{1/2}|B_i\rangle,
\end{eqnarray} 
where the upper and lower signs in the exponents of $T^2$ refer to $y' \in 2\mathbf Z\pm\frac{1}{2}$ respectively.
A particularly interesting choice of the boundary state
$|B_i\rangle$ and $|B_f\rangle$ is the so-called domain-wall initial state $|DW\rangle$, which corresponds to the fermionic state at fixed imaginary time; $|DW\rangle=\prod_{x'<0}c^{\dagger}_{x'}|0\rangle$. Graphically, the domain wall initial state is the edge configuration at fixed semi-integer $y'$ coordinate with all lines thick for negative $x'$  and thin  otherwise. It is easy to realize that choosing these peculiar boundary states at the top and bottom of the strip we can build a square lattice with $N\times N$ sites, coloured in gray in Fig.~\ref{fig_1}, whose sides satisfy DWBC. Taking $x'=n+\frac{1}{2}$ and $y'=m+\frac{1}{2}$, $n-m\in 2\mathbf Z$ and introducing the new integer lattice coordinates $(x,y)$ 
according to
\begin{eqnarray}
\label{xcoord}
x&=&\frac{m+n+(N+1)}{2},\quad 0\leq x\leq N,\\
\label{ycoord}
y&=&\frac{m-n+(N+1)}{2},\quad 1\leq y\leq N,
\end{eqnarray}
we can interpret the eigenvalues of $\rho_{h}(x,y)\equiv\rho_{\rm{even}}(x'(x,y)),y'(x,y))$  to be 1 if the horizontal edge $(x,y)$ is occupied, and 0 otherwise; see Fig.~\ref{fig_1}. 
Thus, in particular, at every horizontal edge $(x,y)$, $0\leq \langle \rho_{h}(x,y)\rangle \leq 1$. The function $\langle\rho_h\rangle$
is equivalent to the polarization considered, for example, in~\cite{BPZ_boundary, syljuaasen}.

 So far, the imaginary-time formalism  is rather general. However, actual calculations can be performed only when the $R$-matrix is the exponential of a quadratic form in the fermionic creation and annihilation operators. Defining~\cite{Baxter}
\begin{equation}
\label{delta}
\Delta=\frac{a^2+b^2-c^2}{2ab},
\end{equation}
this condition restricts the analytical study 
of the six-vertex with the above method to $\Delta=0$: the so-called free fermion point.
At this point, we parameterize $a=c\cos\psi$ and $b=c\sin\psi$ and assume $\psi\in[0,\pi/2]$. In \cite{viti} a
finite-size formula was obtained for $\langle \rho_{\rm{even}}(x',y')\rangle $ in terms of a double integral. An equivalent formula was also before obtained in~\cite{kj} for the density inside the Aztec diamond\footnote{See in particular formula (2.21) in~\cite{kj}, we are grateful to the referee for clarifying this point.}.

 To
allow comparison with Monte Carlo data, it is convenient to expand the double integral in a convergent series of
one-dimensional
integrals\footnote{Notice however that for $m$ close to $\pm N$ the density 
(\ref{density_fz}) is the the product  of an exponentially large function with an exponentially small one.
Therefore a certain care must me taken when handling the numerical integration.} and derive the following expression for the finite-size density of fermions on the horizontal edges
\begin{equation}
\label{density_fz}
\hspace*{-2cm}\langle \rho_h(x,y) \rangle=\sum_{j=0}^{\infty} F^{+}_j F^{-}_j,~~~~~~F^{\pm}_j=\int_{-\pi}^{\pi}\frac{d\kappa}{2\pi}
\sqrt{\frac{\cos\psi-i\sin\psi\tan\kappa}
{\cos^2\psi\cos^2\kappa+\sin^2\kappa}}~e^{\pm i\Phi \pm i(j+\frac{1}{2})\kappa}.
\end{equation}
The phase $\Phi$  is  given by 
\begin{equation}
\hspace*{-1cm}\Phi=k(\kappa)\Bigl(n+\frac{1}{2}\Bigr)+\sqrt{N^2-(m+1/2)^2}~
\frac{e^{\alpha(m)}\varepsilon(-\kappa)
-e^{-\alpha(m)}\varepsilon(\kappa)}{2i},
\end{equation}
and it depends on the lattice coordinates $x$ and $y$ through (\ref{xcoord}) and (\ref{ycoord}).  The functions $\alpha$ and $\varepsilon$ are defined by
\begin{equation}
\alpha(m)={\rm arctanh}\left(\frac{m+1/2}{N}\right),~~\varepsilon(\kappa)=\log\left(\frac{1+\tan\frac{\psi}{2}e^{i\kappa}}{1-\tan\frac{\psi}{2}e^{i\kappa}}\right),
\end{equation}   
and  $k(\kappa)$ is the bijective map $\tan k=\frac{\tan\kappa}{\cos\psi}$. 
 It can be shown that although derived for odd $N$, (\ref{density_fz}) holds also for even $N$.
 In the limit $\psi\rightarrow 0$, the phase $\alpha$ can be eliminated by a change of the contour of integration in (\ref{density_fz}). 
 However, this does not appear to be the case in general. In the scaling limit
 ($x$, $y$, $N\to \infty$ and $x/N$, $y/N$ constant) we can get simpler expressions for the fermionic densities $\langle\rho_h\rangle$ and $\langle\rho_v\rangle$
 on the horizontal and vertical edges. These formulas can be derived by stationary phase techniques; we remand to~\cite{viti} for all the details,  see also the right panel of Fig.~\ref{fig_fs}.  
\section{Numerical Technique}
\label{sec3}
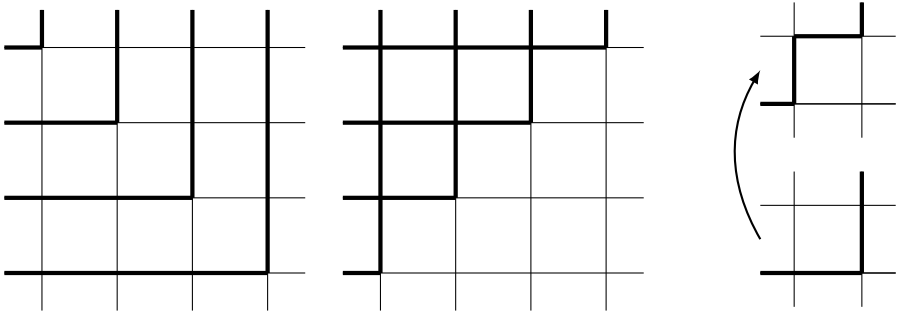
\begin{figure}[t]
\centering
\begin{tikzpicture}[scale=1]
\draw(-0.5,0)--(3.5,0);
\draw(-0.5,1)--(3.5,1);
\draw(-0.5,2)--(3.5,2);
\draw(-0.5,3)--(3.5,3);
\draw(0,-0.5)--(0,3.5);
\draw(1,-0.5)--(1,3.5);
\draw(2,-0.5)--(2,3.5);
\draw(3,-0.5)--(3,3.5);
\draw[ultra thick](0,3.5)--(0,3);
\draw[ultra thick](1,3.5)--(1,2);
\draw[ultra thick](2,3.5)--(2,1);
\draw[ultra thick](3,3.5)--(3,0);
\draw[ultra thick](-0.5,3)--(0,3);
\draw[ultra thick](-0.5,2)--(1,2);
\draw[ultra thick](-0.5,1)--(2,1);
\draw[ultra thick](-0.5,0)--(3,0);
\begin{scope}[xshift=4.5cm]
\draw(-0.5,0)--(3.5,0);
\draw(-0.5,1)--(3.5,1);
\draw(-0.5,2)--(3.5,2);
\draw(-0.5,3)--(3.5,3);
\draw(0,-0.5)--(0,3.5);
\draw(1,-0.5)--(1,3.5);
\draw(2,-0.5)--(2,3.5);
\draw(3,-0.5)--(3,3.5);
\draw[ultra thick] (-0.5,3)--(3,3);
\draw[ultra thick] (-0.5,2)--(2,2);
\draw[ultra thick] (-0.5,1)--(1,1);
\draw[ultra thick] (-0.5,0)--(0,0);
\draw[ultra thick] (0,3.5)--(0,0);
\draw[ultra thick] (1,3.5)--(1,1);
\draw[ultra thick] (2,3.5)--(2,2);
\draw[ultra thick] (3,3.5)--(3,3);
\end{scope}
\begin{scope}[xshift=10cm, scale=0.9]
\draw (-0.5,0)--(1.5,0);
\draw (-0.5,1)--(1.5,1);
\draw (-0.5,0)--(1.5,0);
\draw (0,-0.5)--(0,1.5);
\draw (1,-0.5)--(1,1.5);
\draw[ultra thick] (-0.5,0)--(1,0);
\draw[ultra thick] (1,0)--(1,1.5);
\draw (-0.5,0+2.5)--(1.5,0+2.5);
\draw (-0.5,1+2.5)--(1.5,1+2.5);
\draw (-0.5,0+2.5)--(1.5,0+2.5);
\draw (0,-0.5+2.5)--(0,1.5+2.5);
\draw (1,-0.5+2.5)--(1,1.5+2.5);
\draw[ultra thick] (-0.5,2.5)--(0,2.5);
\draw[ultra thick] (0,2.5)--(0,3.5);
\draw[ultra thick] (0,3.5)--(1,3.5);
\draw[ultra thick] (1,3.5)--(1,4);
\draw[thick, -latex](-0.5,0.5) to [bend left](-0.5,0.5+2.5);
\end{scope}
\end{tikzpicture}
\caption{\textit{Left.} The high state on the left and the low state on the right. The high state contains three vertices flippable up and the low
state three vertices flippable down. \textit{Right.}The flip-up operation. The flip-down operation is obtained by time reversal.}
\label{low}
\end{figure}
Consider the six vertex model with DWBC on a $N\times N$ square lattice as in Fig.~\ref{fig_1}. We may
represent each configuration of the partition function 
 as a collection of $N$ fermionic paths that do not cross
or share an edge, but may intersect at a vertex. For simplicity we will refer to the fermionic paths in imaginary time as \textit{curves}.\footnote{Note that this is 
different from periodic boundary conditions: With periodic boundary conditions, for every number
$s$ in the range $0 \leq s \leq N$, the
partition function contains at least one state (configuration) whose number of curves is exactly $s$.} 
Since the number of curves is fixed, any state may be obtained from any other state by mere perturbation
of curves, and without breaking any curve. The Monte Carlo algorithm  introduced by Allison and Reshetikhin in\cite{reshetikhin}  is based on this observation.  
If instead one wishes to consider periodic boundary conditions, then some other method will have to be implemented, see for example~\cite{barkema} or \cite{syljuaasen}. We now describe in greater detail the numerical technique proposed in~\cite{reshetikhin} and that is used in this paper. 

The local move of such a 
Monte Carlo method is a perturbation of one of the curves seen in  Fig.~\ref{low} at one vertex. We call these perturbations flips. There are two kinds of these;
flip up and flip down. We illustrate a flip up in the right panel of Fig.~\ref{low}. An example of a
flip down is the reverse of the process seen in the figure. Not all vertices are flippable.
The six vertex rule must still be satisfied after the flip. In the high configuration in the left panel of  Fig.~\ref{low}  there are three
vertices flippable up and none flippable down. In the low configuration in Fig.~\ref{low} the situation is the opposite;
three vertices are flippable down and none flippable up. In a configuration there may also be
vertices flippable both up and down.   
With our boundary conditions, the number of curves, $N$, is of course conserved. We may think of
the process of flips as a Markov process on the set of allowed configurations. What follows is
an outline of the algorithm used.

Each flip involves four vertices. 
We call the vertex that gets flipped $ v=(x,y)$; the coordinate $x$ is a half integer with $0<x<N$ and $y$ is an integer $1\leq y\leq N$, see Fig.~\ref{fig_1}. Let us consider
the case where the flip is up. The weight of the local state $S$ is then 
\begin{equation}
 W_v(S) = w(x,y)w(x,y+1)w(x-1,y+1)w(x-1,y)
 \label{wb}
\end{equation}  
before the flip, where each $w(\cdot,\cdot)=a,b,c$ is the weight of a single vertex. After the flip,
the state $S$ has been changed to a new state $S'$ whose weight is 
\be W_v(S') = w'(x,y)w'(x,y+1)w'(x-1,y+1)w'(x-1,y).
\label{wa}
\ee 
It is easy to see that there are sixteen different states $S$. We call the greatest of the
sixteen weights (which of course need not all be distinct) $W_0$. 

To begin the Markov process, we first read a global state from a file. For example, we may begin
with the high state seen in Fig.~\ref{low}. We then determine which vertices are flippable down
only, flippable up only, and flippable both ways. We then randomly generate pairs of vertex coordinates
$(x,y)$ until we find a vertex $v$ which is flippable. 
We then execute the flip with a probability which depends on the state $S'$ after the flip. We must
distinguish between the case where the vertex is flippable only one way and the case where it is
flippable both ways. In the latter case we must also decide which way to flip it. First, let us 
consider the case where there is only one way to flip. We then flip the vertex with probability
\be P=\frac{W_v(S')}{2W_0}.
\label{pflip1}
\ee    
When the vertex is flippable both ways, there are two different local states $S'$ which we call
$S_1'$ and $S_2'$. With probability
\be P_{1,2}=\frac{W_v(S_1')+W_v(S_2')}{2W_0}
\label{pflip2}
\ee
we execute a flip. If it is decided that a flip should be executed, we must decide which one of them
it should be. We flip to state $S_j'$ with probability
   \be P_{j}=\frac{W_v(S_j')}{W_v(S_1')+W_v(S_2')}.
\label{pflip2j}
\ee
If a flip has occurred, then the flippability of the points near the flipped vertex will change. Thus
the lists of flippable vertices must be updated. After this change has been done, we begin to
search for flippable vertices again. We define a Monte Carlo sweep to be $N^2$ vertex flips. 
We continue this process until the system has thermalized. As a practical criterion we assume  thermalization has occurred when
the shape of the boundary of the frozen region does not  
seem to change  any more. A comparison between the arctic curve obtained with Monte Carlo and the exact curves derived in the thermodynamic limit in~\cite{PC, PCZ} will be reported in the next sections. 

Finally we also checked that the integrated autocorrelation time scales very slowly with 
the lattice side-length $N$; that is an indication that we are sampling statistically independent configurations. 
This analysis is reported in the inset on the left panel in Fig.~\ref{fig_fs}, where we plotted the integrated autocorrelation 
time $\tau$ against  $N$ with $\Delta=0$ and $a=b$. The measured quantity is the number of vertices of type $c$, and the time unit 
is $1/8$ of a Monte Carlo sweep.

\section{Disordered regime $|\Delta|\leq 1$}
\begin{figure}[t]
\centering
\begin{tikzpicture}
    \begin{scope}
    \node {\includegraphics[height=5.5cm]{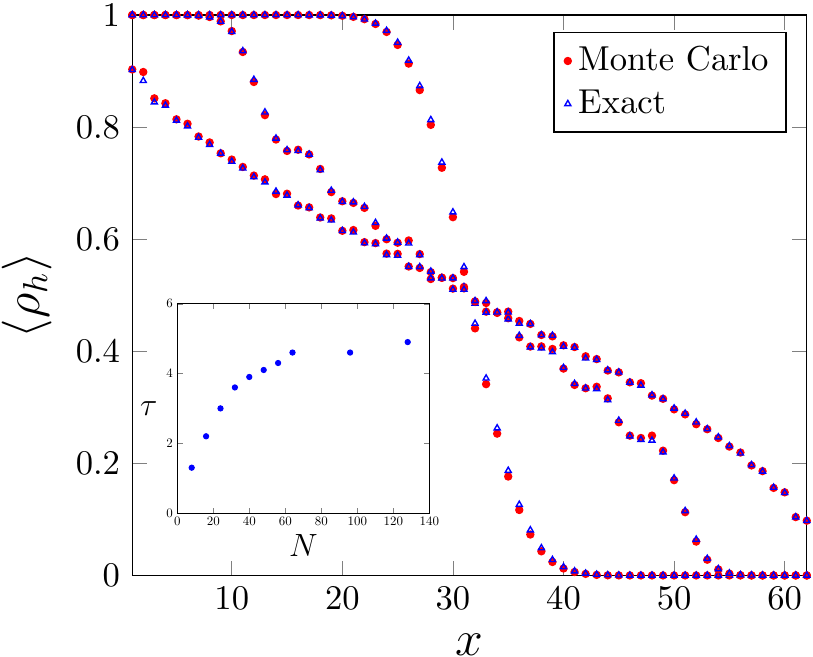}};
    \end{scope}
    \begin{scope}[xshift=6.5cm, yshift=-0.1cm]
    \node {\includegraphics[height=5.5cm]{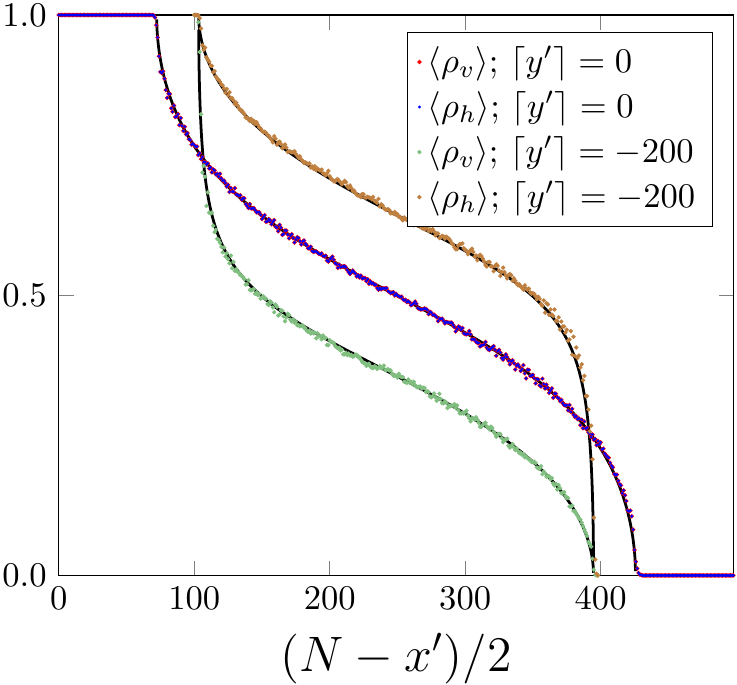}};
    \end{scope}
 \end{tikzpicture}
 \caption{\textit{Left.} From top to bottom the density profiles are plotted along a row at
 constant vertical coordinate $y=1,9,31$; see Fig.~\ref{fig_1}. The lattice as length-side $N=63$, $\Delta=0$ and $a=b$. Inset: Integrated autocorrelation time as a function of the lattice side-length $N$, see Sect.~\ref{sec3} for details. \textit{Right.}  Density profile on horizontal and
 vertical edges along the NW-SE diagonal ($\lceil y' \rceil=0$) and on a direction parallel to it ($\lceil y'\rceil=- 200$). The side-length of the lattice is $N=500$; 
 black solid curves denote analytic
 formulas in~\cite{viti}. When $\lceil y' \rceil=0$, the plots of $\langle \rho_v \rangle $ and $\langle \rho_h \rangle $
 can barely be distinguished. This is expected when $N$ is large \cite{viti}.}
 \label{fig_fs}
\end{figure}

We implemented the algorithm described in the Sec.~\ref{sec3} to study numerically the density profile in the disordered regime.
We begin with a test of the algorithm at $\Delta=0$ where analytic expressions for the density are known.
We further consider the case $\psi=\pi/4$, i.e. $a/c=b/c=1/\sqrt{2}$. On the left panel of Fig.~\ref{fig_fs} is shown a comparison
between Monte Carlo data and (\ref{density_fz}) obtained with $N=63$
and different rows of the lattice $y=1,9,31$: the agreement is excellent. For $N=32$ and $N=64$, similar graphs were plotted in
\cite{syljuaasen}, where the authors used a different Monte Carlo algorithm and an unpublished integral representation of the
exact expectation value. Notice that for finite $N$, the density shows plateaux whose
number increases as we move towards the center of the lattice. The presence of such a staircase pattern 
is related to the fact that moving from left to right the density decreases by a finite amount when crossing a fermionic trajectory.
Moreover, since there are $N$ particles travelling inside the gray region in Fig.~\ref{fig_1}, this effect is more pronounced for smaller lattices.
The same phenomenon happens in real-time evolution, as discussed for example in~\cite{BG, Eisler, current}.
In the thermodynamic limit $N\rightarrow\infty$, the density profile becomes smoother as can be seen on the right panel of Fig.~\ref{fig_fs},
where we considered a lattice with side-length
$N=500$.
In particular, we show a comparison between the asymptotic formulas obtained in~\cite{viti} for the densities
on horizontal and vertical edges  $\langle\rho_h\rangle$ and $\langle\rho_v\rangle$  along the NW-SE
diagonal in $(x,y)$ coordinates (or $\lceil y' \rceil=0$) and parallel to the same diagonal ($\lceil y'\rceil=- 200$).
The agreement is again excellent: exact curves are plotted in black and barely visible.
The tails of the density profiles in this case are
described by the Airy kernel~\cite{kj2000, kj},
see also~\cite{FS}, but unfortunately the  numerical resolution is not good enough to allow a quantitative study here. We plan to come back on this
problem in the future.
\begin{figure}[t]
\centering
\begin{tikzpicture}
    \begin{scope}
    \node {\includegraphics[height=5.5cm]{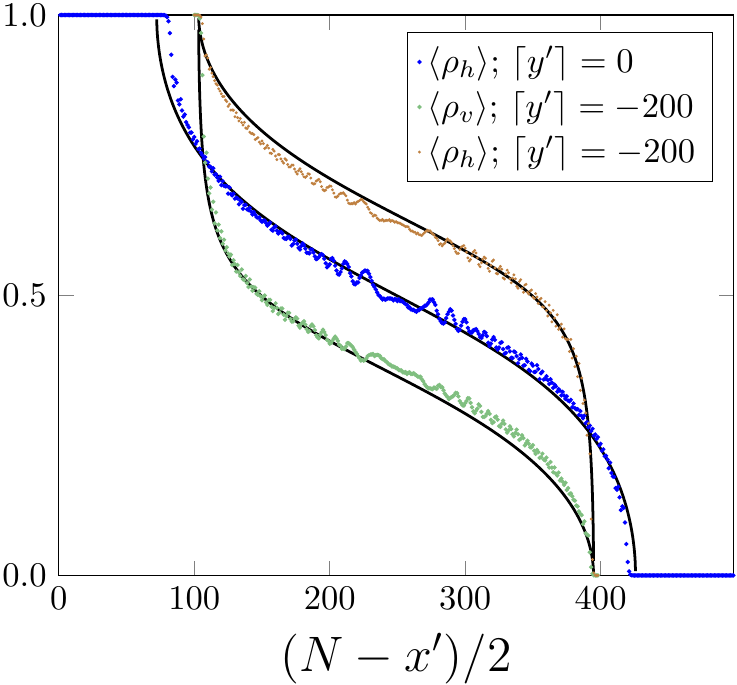}};
    \end{scope}
    \begin{scope}[xshift=6.5cm, yshift=0.3cm]
    \node {\includegraphics[height=4.9cm]{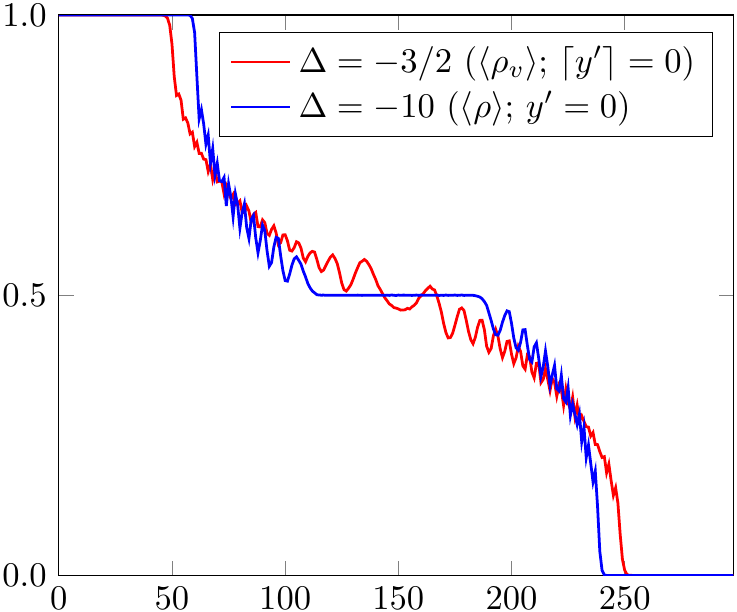}};
    \end{scope}
 \end{tikzpicture}
 \caption{ \textit{Left.} Various density profiles at $\Delta=-1$ for $N=500$ and 
 $a=b$ as discussed in the main text. For comparison, the same black solid curves as in the right panel of Fig.~\ref{fig_fs} are shown.
 \textit{Right.} In red, the expectation value of the density on vertical edges
  when $\Delta=-3/2$ and $a=b$. The measurement is done diagonally with $y'=-1/2$. 
 The points have been joined by a curve for graphical reasons. In blue, the expectation value of the density on vertices at $\Delta=-10$ and $a=b$.}
 \label{fig_graph}
\end{figure}

 To obtain the curves on the right panel of Fig.~\ref{fig_fs}, we ran the code for $10^5$ Monte Carlo sweeps to achieve thermalization, and then measured the density after each of the following $35000$ Monte Carlo sweeps. For a lattice with side-length $N=500$  the whole process might take approx. 45h of CPU time. We did not investigate in detail the dependence of the thermalization time from the size of the lattice but expectation is that it will increase polynomially~\cite{reshetikhin}.
 
Similar plots of the density profiles can be obtained in the whole disordered regime. As an example, in the left panel 
of Fig.~\ref{fig_graph} we show the expectation value of the density on horizontal edges, $\langle \rho_h \rangle$, on the NW-SE diagonal ($\lceil y'\rceil=0$) at 
$\Delta=-1$ $(a=b)$. We also show the expectation value of the density on the vertical and horizontal edges at constant $\lceil y' \rceil=-200$. In the plot solid black 
curves denote again the exact asymptotic formula at $\Delta=0$, as in Fig.~\ref{fig_fs}. Now, of course, these curves are included only for comparison. 
Numerical results at $\Delta\not=0$ could be  affected by finite size effects in a different way with respect to $\Delta=0$ and this type of correction is largely unknown. In any case our numerical results at $\Delta=-1$ show now a more pronounced oscillatory pattern, see in particular the blue curve in Fig.~\ref{fig_graph} left. Unfortunately at present an analytic understanding of this phenomenon is not available.  
  Finally we tested the average arctic curve obtained from Monte Carlo against its exact expression conjectured in \cite{PC} at 
  both $\Delta=-\frac{1}{2}$ and $\Delta=-1$ for equal weights $a=b$. The results are in the left panel of 
  Fig.~\ref{fig_shape} and show impressive agreement already at $N=500$. In particular the first two pictures on the 
  left show typical lattice configurations  for these values of $\Delta$. In  the other two next to them is plotted 
  $2\langle\rho\rangle-1$, where $\rho$ is the density on a vertex; vertex $a_1$ has density $\rho=1$, $a_2$ has density 0, and the remaining four vertices have density $1/2$.
  In the SW corner of the lattice the boundary conditions force $\langle\rho\rangle=1/2$. The blue continuous curve is the exact
  arctic curve in~\cite{PC} and the red one is the arctic circle, shown for comparison. To our best knowledge a comparison between the analytic expression
  for the limiting shapes and their numerical estimate was performed only at $\Delta=0$ and
  at the ice-point $\Delta=\frac{1}{2}$, exploiting the \textit{Coupling From the Past} sampling method~\cite{PW}. 

\section{Antiferromagnetic regime $\Delta<-1$}
\begin{figure}[t]
\centering
\includegraphics[height=6.5cm]{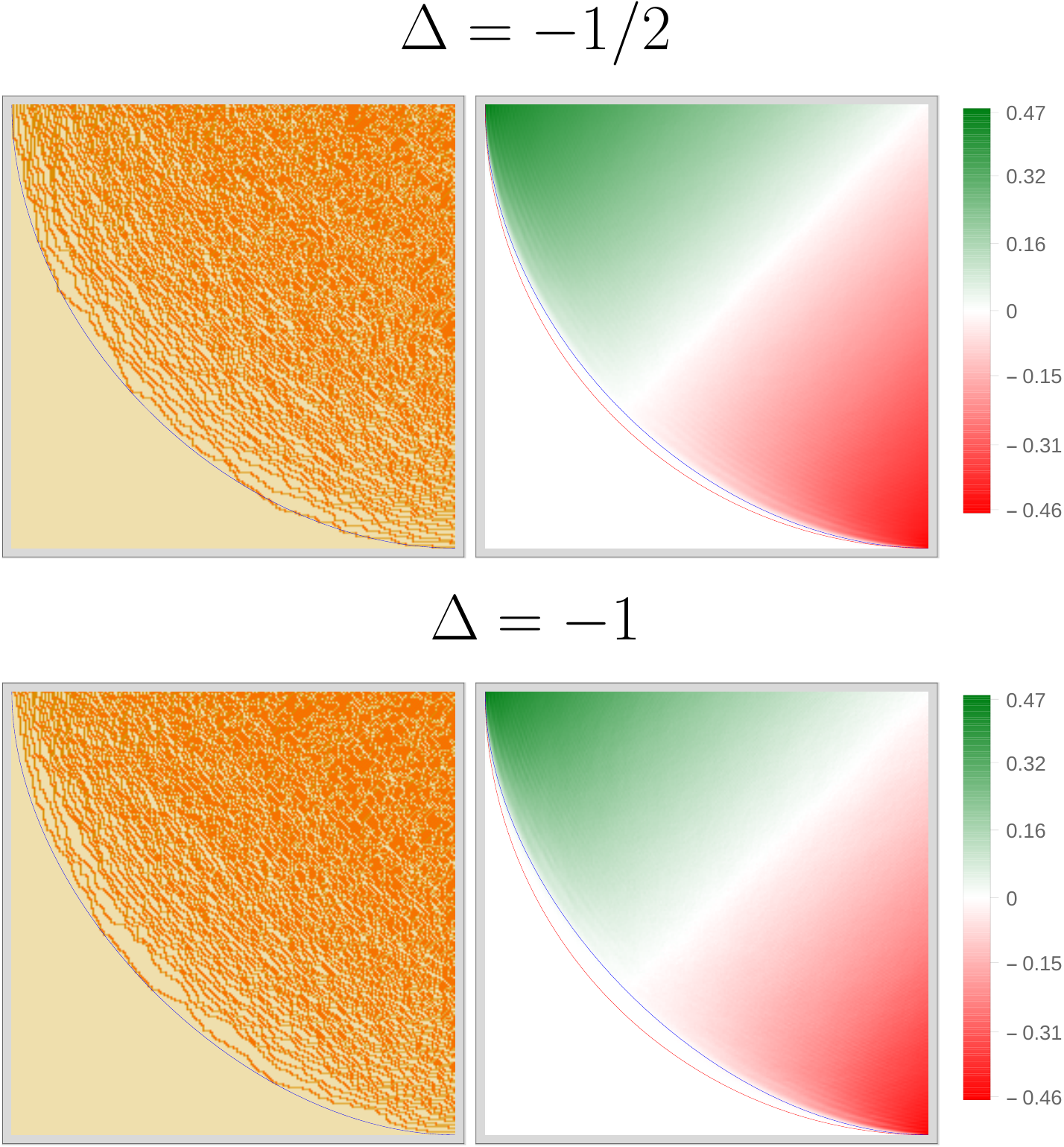}\quad
\includegraphics[height=6.5cm]{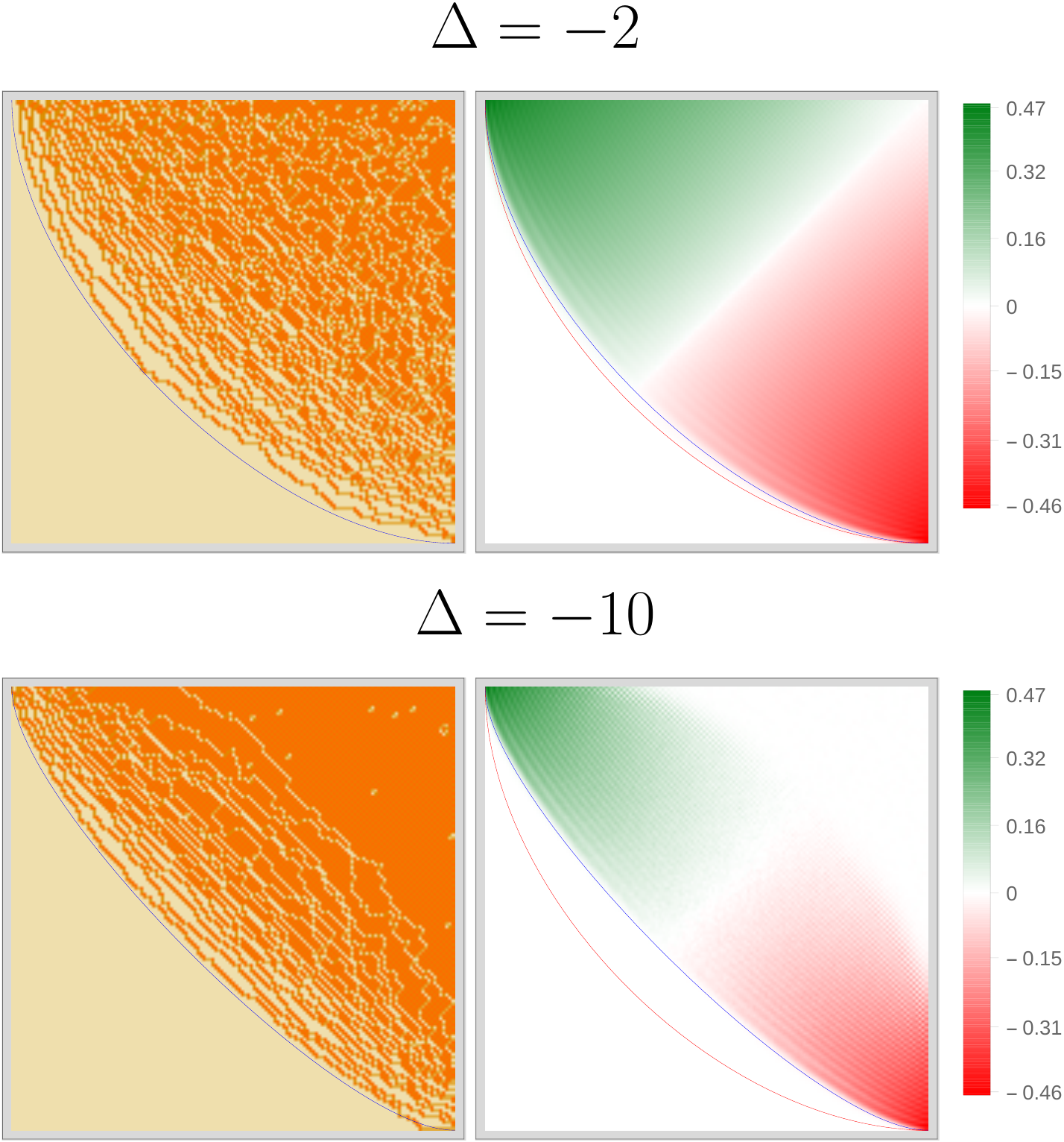}
 \caption{
\textit{Left.} Typical configurations in the disordered regime  at $\Delta=-1/2$ on top and $\Delta=-1$ at the bottom, 
 both for $a=b$ and $N=500$. On the right we plot $\langle 2\rho\rangle-1$ in the SW corner, blue curves are analytic predictions 
 in~\cite{PC} and the red curve is the arctic circle.  
 \textit{Right.}  Typical configurations in the antiferromagnetic regime  at $\Delta=-2$ on top and $\Delta=-10$ at the bottom, 
 both for $a=b$ and $N=300$. On the right we plot $\langle 2\rho\rangle-1$ in the SW corner, blue curves are analytic predictions 
 in~\cite{PCZ} and the red curve is the arctic curve at $\Delta=-1$.}
\label{fig_shape}
\end{figure}
In the antiferromagnetic regime, an antiferromagnetic bubble will form in the middle of the lattice when $\Delta$ is sufficiently
below $-1$ \cite{reshetikhin}. It  consist of alternating $c_1$ and $c_2$ vertices, and it is therefore doubly degenerate. In the limit $\Delta\rightarrow-\infty$ this antiferromagnetic island will assume a rectangular shape~\cite{syljuaasen, reshetikhin}. The state  
inside the bubble may shift over time during the Monte Carlo simulation, and this shift takes longer time the larger
the bubble is; that is, the further $\Delta$ is below $-1$. For this reason, it may take a very long time to accurately
measure the expectation value of the density on edges. If $\Delta$ is far below $-1$, such a measurement may impractical
 with our Monte Carlo algorithm. However, we think that the measurement of the expectation value of the density on 
vertices is still accurate, since $\rho$ is the same for both antiferromagnetic states.  

\begin{figure}[t]
\centering
\includegraphics[height=6cm]{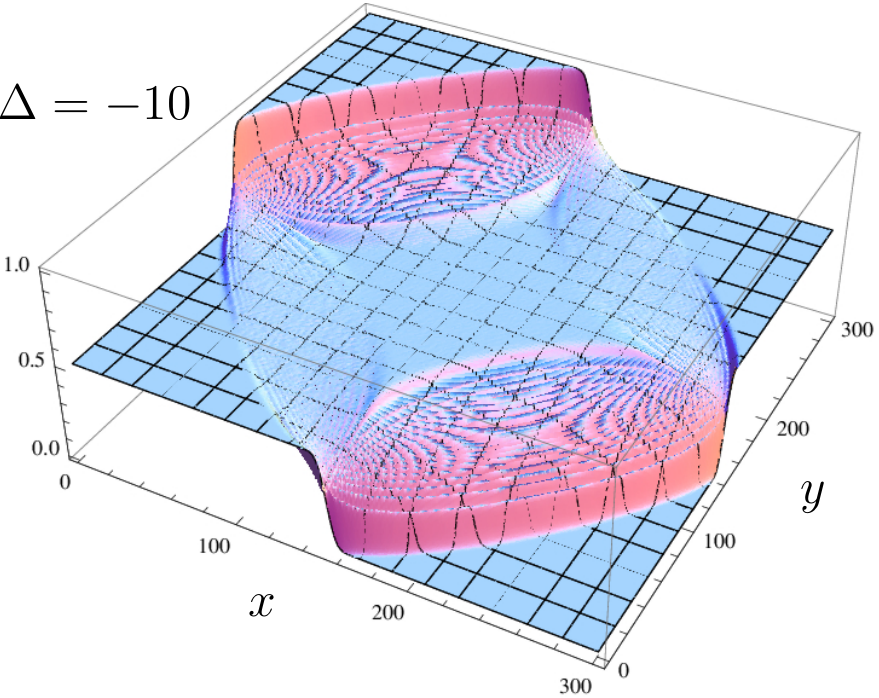}
 \caption{The expectation value of the density on vertices, $\langle \rho\rangle $, when $\Delta=-10$ and $a=b$.
 The coexistence of three different phases is clearly seen. The antiferromagnetic region in the middle
is surrounded by a disordered region, which in its turn is surrounded by a ferromagnetic region.}
\label{fig3d}
\end{figure}

In the right panel of Fig.~\ref{fig_graph} we have plotted $\langle \rho_v \rangle $ for $\Delta=-3/2$ and 
$\langle \rho \rangle $ for $\Delta=-10$. 
Both values of $\Delta$ pertain to $N=300$. The vertex density $\langle \rho \rangle$
is measured along the main diagonal from NW to SE ($y'=0$), and $\langle \rho_v \rangle $ is measured on vertical edges one half
step below the same diagonal. If we compare the right panel of Fig.~\ref{fig_fs} with the left panel of Fig.~\ref{fig_graph}, we observe that oscillations are more pronounced 
in the antiferromagnetic regime than when $|\Delta|\leq 1$; similar findings are discussed in~\cite{Dutch} with a different algorithm. Furthermore the existence of a non-critical antiferromagnetic phase in the center of the
lattice is also confirmed by the emergence of a  flat central region with constant $\langle\rho\rangle=1/2$.  In the right panel of Fig.~\ref{fig_shape}, we have gathered two typical lattice configurations; at $\Delta=-2$ and $\Delta=-10$ respectively. The mean value 2$\langle \rho \rangle-1$ is also included. The blue curve is the conjectured arctic curve
 in ~\cite{PCZ} and the red one is the arctic curve at $\Delta=-1$, shown for comparison. The agreement between the limiting shape and
 the one obtained from the Monte Carlo data is excellent. Looking at the pictures with $\Delta=-10$ in the right panel of Fig.~\ref{fig_shape}, we also see clearly that there is another curve which separates the disordered region from the antiferromagnetic region in the center. This curve has four cusps and can be clearly identified in the three dimensional density plot with $\Delta=-10$ and $N=300$ in Fig.~\ref{fig3d}. In the thermodynamic limit $N\rightarrow\infty$ three phases can coexist: a frozen phase in each of the four corners, surrounding a disordered inhomogeneous phase which surrounds a non-critical antiferromagnetic droplet. This antiferromagnetic region at the center is dominated by vertices of type $c$. At present no analytic understanding of  the phenomenon of separation is available. It is worth mentioning that an analogous coexistence of three phases can also be found  in a free fermionic model with staggered magnetic field, when time evolution is performed in imaginary time  from a domain-wall initial state, see for example Fig.~15 in~\cite{viti}. In the limit of strong magnetic field the gapped region with N\'eel order also appears to have a rectangular shape~\cite{Stephan}.
 
Finally, the presence of such an antiferromagnetic bubble for large negative values of $\Delta$ can be argued from a quantum mechanical point of view. We recall that the in the so-called Hamiltonian limit~\cite{Baxter, Thacker} the diagonal-to-diagonal transfer matrix of the
 six-vertex model described in Sec.~\ref{sec_1} can be written as $\tilde{T}=1+\varepsilon H_{XXZ} +O(\varepsilon^2)$, where $H_{XXZ}$ is the $XXZ$ Hamiltonian and $\varepsilon$ a small parameter; explicitly  
\begin{equation}
\label{XXZ}
H_{XXZ}=-\sum_{i=1}^{L} \left(\sigma_i^x\sigma_{i+1}^{x}+\sigma_i^y\sigma_{i+1}^{y}+\Delta\sigma_i^z\sigma_{i+1}^{z}\right).
\end{equation} 
Here $L$ is the total length of the chain (the strip in Fig. \ref{fig_1}) that is taken to be very large, so that the boundary conditions at the two ends  are unimportant. From now on, we shall assume that the ends are free. The Hamiltonian (\ref{XXZ}) conserves the total magnetization $M=\sum_{i}\sigma_i^z$ and
it has been tacitly assumed that $M=0$. The eigenvectors of the quantum Hamiltonian describe a typical configuration along a row of the statistical mechanical model, i.e. a line at constant imaginary time $y'$ in Fig.~\ref{fig_1}.

 For $\Delta<-1$  the chain  has a gapped ground state with N\'eel ordering; in the limit $\Delta\rightarrow-\infty$ the ground state is either  the N\'eel or the anti-N\'eel state. As we have discussed in Sec.~\ref{sec_1}, DWBC in the six vertex model can be obtained by evolution in imaginary time, starting from a domain wall initial state. In the spin basis where $\sigma_i^z$ is diagonal, we have $|DW\rangle=|\uparrow\uparrow\uparrow\dots\uparrow\downarrow\downarrow\downarrow\dots\downarrow\rangle$. The  state of the chain after imaginary time $T$ (in units of $\varepsilon$) is obtained as $|DW(T)\rangle=e^{-TH_{XXZ}}|DW\rangle$. For large imaginary time the initial state is projected into the lower energy state with the same magnetization, in the case of the domain wall initial state this is the gapped antiferromagnetic ground state. This simple argument shows that far apart from the upper and lower boundaries of the strip in Fig.~\ref{fig_1}, lattice configurations in the Hamiltonian limit should manifest an antiferromagnetic order, at least near the center.

\section{Conclusions and Perspectives}
In this paper we have implemented the algorithm of Allison and Reshetikhin to study density profiles in the six-vertex model with DWBC, varying the control parameter $\Delta$. We have also performed several  tests of the reliability of the Monte Carlo algorithm. In particular we have compared  data for the density at the free fermion point with exact results derived in the thermodynamic limit in~\cite{viti} and with an explicit finite size formula presented in Sec.~\ref{sec_1}. We have
also showed that the arctic curves extracted from the numerics are in excellent agreement with  the curves conjectured in~\cite{PC, PCZ} both in the disordered and anti-ferromagnetic regime. This analysis suggests that the algorithm in~\cite{reshetikhin} can be used for quantitative purposes, although in the antiferromagnetic regime it might be impractical due to the presence of two degenerate ground states  where the Markov chain could be trapped. 

 We have then confirmed the coexistence of three different phases in the antiferromagnetic regime and showed that the density profiles present in this case a flat central region and an inhomogeneous part with more pronounced oscillations. It would be interesting
 to analyze whether also  a Tracy Widom distribution at the edges of the profile arises as argued in~\cite{PC} for the whole disordered regime. 

 Many  generalizations are possible and are currently under investigation by the authors. Among them we mention the study of the phase diagram for other type of boundary conditions like alternating boundary conditions~\cite{TRK1, TRK2} or partial domain wall boundary conditions~\cite{FW, BL}. The algorithm can also easily be adapted to lattices of different shape, such as an L-shape~\cite{CP_L}.

\section*{Acknowledgements} We are grateful to G. Gori, J. Lamers, A. Pronko, N. Reshetikhin and J-M. St\'ephan for correspondence and discussions. We thank in particular F. Colomo for suggestions, criticism and a careful reading of the manuscript.


\section*{References}
\begin{thebibliography}{999}

\bibitem{P35} L. Pauling, The Structure and Entropy of Ice and of Other Crystals with some Randomness of Atomic Arrangement, J. Am.
 Chem. Soc. 57 (12): 2680-2684, (1935).

\bibitem{Lieb} E. Lieb, Phys. Rev. 162, 162, (1967); E. H. Lieb, Phys. Rev. Lett. 18, 1046, (1967); 19, 108, (1967). 

\bibitem{Baxter} R. Baxter, Exactly Solved Models in Statistical Mechanics, Academic Press, San Diego (1987).

\bibitem{zj_bc} P. Zinn Justin, The Influence of Boundary Conditions in the Six-Vertex Model, ArXiv: cond-mat/0205192 (2002).

\bibitem{kzj} V. Korepin and P. Zinn-Justin. Thermodynamic limit of the 
six-vertex model with domain wall boundary conditions. 
J. Phys. A 33 (2000), p. 7053-7066; Inhomogeneous Six Vertex model with Domain Wall Boundary Conditions and Bethe Ansatz, J. Math. Phys. 43, 3261-3267, (2002).

\bibitem{Korepin} V. Korepin, Calculation of Norms of Bethe Wavefunctions, Comm. Math. Phys. 86, 391-418 (1982).

\bibitem{KBI} V. Korepin, N. Bogoliubov and A. Izergin,
Quantum inverse scattering method and correlation
functions, Cambridge University Press (1993).

\bibitem{Izergin} A. Izergin, Partition Function of the 6-Vertex Model in a Finite Volume, (Russian) Dokl. Akad. Nauk USRR, v. 297, 331-333, (1987).

\bibitem{ICK} A. Izergin, D. Coker, and V. Korepin,
Determinant formula for the six-vertex model, J. Phys. A 25 (1992), 4315–4334.

\bibitem{Bleher} P. Bleher and V. Fokin, Exact Solutions of the Six Vertex Model
 with Domain Wall Boundary Conditions. Disordered Phase. Comm. Math. Phys. 1,
  223-284 (2006); P. Bleher and K. Liechty, Exact Solutions of the Six Vertex Model
 with Domain Wall Boundary Conditions. Ferroeletric Phase.
  Comm. Math. Phys. 286: 777 (2009); P. Bleher and K. Liechty, Exact Solutions of the Six Vertex Model
 with Domain Wall Boundary Conditions. Antiferroeletric Phase.
  Comm. Math. Phys. 286: 777 (2009).


\bibitem{zj2000} P. Zinn-Justin, Six-vertex model with domain wall boundary conditions and one-matrix model, Phys. Rev. E 62 (2000), 3411–3418.

\bibitem{syljuaasen} O. F. Sylju\aa sen and M. B. Zvonarev.  
Directed-loop Monte Carlo simulations of vertex models.
Phys. Rev. E 70 (2004), 016118. 

\bibitem{KO} R. Kenyon and A. Okounkov, Limit shapes and the complex Burgers equation, Acta Math. 199 (2007), 263–302.

\bibitem{KOS} R. Kenyon, A. Okounkov, and S. Sheffield, Dimers and amoebae, Ann. of Math. 163 (2006), 1019–1056.

\bibitem{Propp} W. Jockush, J. Propp and P. Shor, Random Domino Tilings and the Arctic Circle Theorem, Arxiv: math.CO/9801068

\bibitem{FS} P. Ferrari and H. Spohn, Domino Tilings and the six-vertex model at its free fermion point, J. Phys. A: Math. Gen. 39 (2006), 10297-10306.

\bibitem{ZJ_rev} P. Zinn-Justin, Six-vertex, loops and tiling models: integrability and combinatorics, ArXiv:0901.0665. 

\bibitem{BPZ_boundary} N. Bogoliubov, A. Pronko and M. Zvonarev, Boundary correlation functions of the six-vertex model, J. Phys. A Math. Gen. 35, 5525-5541 (2002).

\bibitem{viti} N. Allegra, J. Dubail, J-M. St\'ephan, and J. Viti. 
Inhomogeneous field theory inside the arctic circle. J. Stat. Mech. (2016) 053108.

\bibitem{PC} F. Colomo and A. Pronko, The arctic curve in the six vertex model, J. Stat. Phys. 138, 662-700 (2010).

\bibitem{PCZ} F. Colomo, A. Pronko and P. Zinn Justin, The arctic curve of the domain wall six vertex model in its anitferroelectric regime, J. Stat. Mech: Theor. Exp. (2010) L03002.

\bibitem{CS} F. Colomo and A.Sportiello,  Arctic curves in the six-vertex model on generic domains: The Tangent Method, J. Phys. A 164 (2016) 1488.

\bibitem{Abanov} A. Abanov, Hydrodynamics of correlated systems. Emptiness Formation Probability and Random Matrices,
 \lq \lq Applications of Random Matrices in Physics\rq \rq, Les Houches Summer School,  (2004).   
 
\bibitem{Reshetikhin2} N. Reshetikhin, A. Sridhar, Integrability of Limiting Shapes of the Six Vertex Model,  ArXiv:1510.01053 (2015).  
 
\bibitem{reshetikhin} D. Allison and N. Reshetikhin.  
Numerical study of the 6-vertex model with domain wall boundary conditions.
Annales de l'Institut Fourier, Tome 55, n$^o$ 6 (2005), p. 1847-1869.




\bibitem{barkema} G. T. Barkema and M. E. J. Newman. Monte Carlo simulation of ice models. Phys. Rev. E 57 (1998), 1155.

\bibitem{cugliandolo} L. Cugliandolo, G. Gonnella and A. Pelizzola, Six vertex model with Domain Wall Boundary Conditions in the Bethe-Peierls approximation, J. Stat. Mech, P06008 (2015).


\bibitem{kj2000} K. Johansson. Shape Fluctuations and Random Matrices. 
Communications in Mathematical Physics (2000), Vol. 209, p. 437-476.  

\bibitem{kj} K. Johansson. The arctic circle boundary and the Airy process. The Annals 
of Probability (2005), Vol. 33, No. 1, p. 1-30.  

\bibitem{PS} M. Praehofer and H. Spohn, Scale Invariance of the PNG Droplet and the Airy Process,  J. Stat. Phys. 108 (5-6): 1071-1106 (2002).

\bibitem{TW} H. Widom and C. Tracy, Level Spacing distributions and the Airy kernel, Phys. Lett. B, 305, 115-118 (1994).


\bibitem{resh_rev} N. Reshetikhin,
 Lectures on integrable models in statistical mechanics, 
 In: \lq\lq Exact methods in low-dimensional statistical physics and
   quantum computing\rq\rq, Proceedings of Les Houches School in Theoretical Physics, Oxford University Press, (2010).

\bibitem{BG} E. Bettelheim and L. Glazman, Quantum ripples over a semi-classical shock, Phys. Rev. Lett.  109, 260602 (2012). 

\bibitem{Eisler} V. Eisler and Z. Racz, Full counting statistics in a propagating front and random matrix spectra, Phys. Rev. Lett. 110, 060602 (2013).

\bibitem{current} J. Viti, J-M. St\'ephan, J. Dubail and M. Haque, Inhomogeneous quenches in a free fermionic chain: exact results,  EPL 115 (2016) 053108.

\bibitem{PW}  J. Propp and D. Wilson, Exact sampling with coupled Markov chains and applications to statistical mechanics, Random Struct. Algor. 9 (1996), 223–252.

\bibitem{Dutch} R. Keesman and J. Lamers, A numerical study of the F-model with domain-wall boundaries, arXiv:1702.05474 (2017).

\bibitem{Stephan} J-M. St\'ephan, private communication.

\bibitem{Thacker} H. Thacker, Exact Integrability in quantum field theory and statistical systems, Rev. Mod. Phys. 53 2, 1981.


\bibitem{TRK1} T. Tavares, G. Ribeiro, V. Korepin, The entropy of the six-vertex model with variety of different boundary conditions, J. Stat. Mech. (2015) P06016.

\bibitem{TRK2} T. Tavares, G. Ribeiro, V. Korepin, Influence of boundary conditions on bulk properties of six-vertex model, J. Phys. A: Math. Theor. 48 (2015) 454004.    

\bibitem{FW} O. Foda, M. Wheeler, Partial domain wall partition functions, Journal of High Energy Physics, 2012, Volume 2012, Number 7, 186.

\bibitem{BL} P. Bleher and K. Liechty, Six-vertex model with partial domain wall boundary conditions: ferroelectric phase,  	J. Math. Phys. 56 (2015) 023302


\bibitem{CP_L} F. Colomo and A. Pronko, Thermodynamics of the six-vertex model in an L-shaped domain, Comm. Math. Phys. 339 (2015), 699-728. 


\end{thebibliography}
\end{document}